\documentclass{elsart}
\usepackage{amsmath}
\usepackage{amssymb}
\usepackage{epsfig}
 \usepackage{epsf}
 \usepackage{epsfig}
 \usepackage{color}
    \newcommand{\ba}{\begin{eqnarray}}
    \newcommand{\ea}{\end{eqnarray}}
    \newcommand{\be}{\begin{equation}}
    \newcommand{\ee}{\end{equation}}

    \newcommand {\bp} {{\mathbf p}}

\newcommand {\fslash}[1]{{#1\kern -0.45em /\kern 0.3em}}

\begin{document}
\runauthor{PKU}
\begin{frontmatter}

\title{Massive Overlap Fermions on Anisotropic Lattices\thanksref{fund}}


\author[PKU]{Xin Li},
\author[PKU]{Guozhan Meng},
\author[PKU]{Xu Feng},
\author[PKU]{Chuan Liu}

\address[PKU]{School of Physics, Peking University\\
              Beijing, 100871, P.~R.~China}

\thanks[fund]{This work is supported by the Key Project of National Natural
Science Foundation of China (NSFC) under grant No. 10235040, No.
10421003, No. 10675005 and supported by the Trans-century fund and
the Key Grant Project of Chinese Ministry of Education (No.
305001).}

\begin{abstract}
 We formulate the massive overlap
 fermions on anisotropic lattices.
 We find that the dispersion relation for the overlap
 fermion resembles the continuum form
 in the low-momentum region once the
 bare parameters are properly tuned.
 The quark self-energy
 and the quark field renormalization constants are
 calculated to one-loop in bare lattice perturbation theory.
 Using tadpole improved perturbation theory, we
 establish the relation between the well-tuned
 parameters and the quark mass once the bare coupling
 is given. We argue that massive domain wall quarks
 might be helpful in lattice QCD studies on heavy-light
 hadron spectroscopy.
\end{abstract}
\end{frontmatter}


\section{Introduction}

 It is known that chiral symmetry plays an essential role
 in low-energy Quantum Chromodynamics (QCD), which
 is believed to be the underlying fundamental theory for strong interactions.
 Another predominant feature for low-energy QCD is its non-perturbative
 nature. Lattice QCD provides a genuine non-perturbative theoretical framework for the
 study of low-energy QCD from first principles.
 Chirality is associated with masslessness of fermions.
 Incorporating massless fermions have always been a great challenge
 in lattice studies due to both theoretical and numerical difficulties.
 In recent years, considerable progress has been made in
 understanding chiral symmetry on the lattice. Domain wall
 fermions~\cite{kaplan92:DWF_idea,shamir93:DWF_boundary} and
 the overlap fermions~\cite{neuberger93:overlap_prl,neuberger94:overlap_npb,neuberger98:massless_vector,neuberger98:massless_quark,neuberger98:more_massless_quark}
 have emerged as two new candidates in the formulation of lattice fermions which
 have much better chiral properties than the conventional
 Wilson or staggered fermions. Since chiral symmetry
 is so crucial to the theory of QCD, it is therefore desirable
 to use these new fermions if possible.

 On the other hand, anisotropic lattices have been used extensively
 on heavy hadronic states and they proved to be extremely
 helpful in various applications. These include: glueball
 spectrum calculations~\cite{colin99,chuan01:gluea,chen06:glueball},
 charmonium spectrum calculations~\cite{chen01:aniso,CPPACS02:aniso},
 charmed meson and charmed baryon calculations~\cite{juettner03:Ds,lewis01:aniso}
 and hadron-hadron scattering
 calculations~\cite{chuan02:pipiI2,chuan04:KN,chuan04:Kpi,chuan04:pipi}.
 Note that many of the above mentioned studies involve
 light quarks for which chiral symmetry is essential.
 It is therefore desirable to
 use either domain wall or overlap fermions which have
 better chiral properties.
 Indeed, much of the systematic uncertainties in these
 studies originates from chiral extrapolations.
 It is therefore tempting
 to study the new lattice fermions (domain wall fermions or overlap fermions)
 on four-dimensional anisotropic lattices. In a previous study, we
 have formulated domain wall fermion on anisotropic lattices using
 bare perturbation theory to one-loop order~\cite{chuan06:DWF_perturb}.
 In this paper, we will
 perform a similar study for the overlap fermions.
 These studies provide us with some guiding information
 on the tuning of the parameters in the corresponding fermion action
 which is necessary in realistic Monte Carlo simulations.

 In this paper, we study massive overlap fermions
 on anisotropic four-dimensional lattices.
 For the gauge action, we adopt
 the tadpole improved gauge actions~\cite{colin97,colin99} which have
 been used in various lattice calculations.
 The fermion action on anisotropic lattices generally contains
 more parameters than its isotropic counterparts. These
 parameters have to be tuned properly in order to yield a correct
 continuum limit. We will first address this issue in the case
 of free overlap fermions on anisotropic lattices. It is found that,
 in order to restore the normal relativistic dispersion relation for
 the quark, parameters of the fermion action have to be tuned accordingly.
 Then, we compute the quark propagator in lattice perturbation theory
 to one-loop. Quark field and quark mass  renormalization
 constants are obtained for various values of the bare parameters.
 This perturbative calculation
 serves as a guidance for further non-perturbative Monte Carlo simulations.
 Similar perturbative calculations have been performed in the case of isotropic
 lattice~\cite{yamada98:overlap_perturb,yamada99:overlap_perturb,ishibashi00:overlap_perturb,fujikawa02:overlap_perturb}.
 Our calculation is an extension of these to the case of anisotropic lattice.
 The use of both anisotropic lattices and overlap fermions to treat
 relativistic heavy quarks on the lattice was first advocated in
 Ref.~\cite{kfliu02:overlap_aniso} where the authors have considered
 the dispersion relation of quarks in a quenched numerical study.

 This paper is organized as follows. In section 2,
 overlap fermion action on anisotropic lattices is given.
 In section 3, the free overlap fermion propagator on anisotropic lattices
 is presented and the dispersion relation of the free overlap fermion is studied.
 It is found that hopping parameters
 of the fermion action have to be tuned properly, according
 to the value of the bare quark mass, so as to have the correct
 continuum limit for the massive quark.
 In section 4, the
 calculations of fermion self-energy to one-loop
 is presented. Numerical results for
 the renormalization factors for the quark field and
 the current quark mass are listed.
 In section 5, we will conclude with some remarks and outlook.
 The one-gluon and two-gluon vertex functions are listed in the appendix.

 \section{The overlap fermions on anisotropic lattices}
 \label{sec:action}

 The fermion action for the exact massless overlap quark has the following
 form~\cite{neuberger98:massless_vector,neuberger98:massless_quark,neuberger98:more_massless_quark}:
 \be
 S_F = \sum_{x,y} \bar{\psi}(x) \, D_{xy} \, \psi(y),
 \ee
 with the (massless) overlap Dirac operator given by:
 \be
 \label{eq:overlapD_massless}
  D(m=0)= \left( 1+X\frac{1}{\sqrt{X^\dagger X}}\right).
 \ee
 The operator $X$ appearing in the above equation
 is the Wilson-Dirac operator on anisotropic lattices which
 we write as:
 \ba
 X_{xy}&=&\frac{1}{2}\sum_{\mu =1}^{4} \kappa_\mu \left[ \gamma_\mu\left\{\delta_{x+\hat
 \mu, y}
 U_\mu(x) - \delta_{x, y+\hat\mu}U_\mu^\dagger(y)\right\} \right.\nonumber\\
 &&\left. + r_\mu\left\{2\delta_{x,y}-\delta_{x+\mu, y} U_\mu(x) - \delta_{x,
 y+\mu}U_\mu^\dagger(y)\right\}\right] + M_0
 \delta_{x,y},\nonumber\\
 \ea
 where $r_\mu$ represents Wilson parameters introduced to remove the doublers
 from low-energy spectrum.
 In the case of anisotropic lattice, we will also use the
 convention: $r_0=r_t$, $r_i=r_s$.
 The parameter $M_0$ is usually taken to be negative
 so as to restore chiral properties of the quark.
 Since one can scale the operator $X$ by a constant factor
 without changing fermion matrix $D$, one might just fix
 the parameter $\kappa_s$ to some constant, say $+1$.
 We will also set the temporal Wilson parameter $r_t=1$.
 It is well-known that the massless overlap
 Dirac operator $D(m=0)$ satisfies the Ginsparg-Wilson
 relation~\cite{ginsparg-wilson:1982}
 which can be viewed as a generalization of the continuum
 chiral symmetry to the lattice~\cite{luscher98:GWL_symmetry,hasenfratz98:GW_symmetry}.
 The Dirac operator also has a massless mode which has definite chirality.

 For the massive quarks, one has to modify the massless overlap
 Dirac operator given in~(\ref{eq:overlapD_massless})
 to~\cite{neuberger98:massless_vector}:
 \be
 \label{eq:overlapD_massive}
 D(m) = \left( 1+\frac{m}{2} +\left(1-\frac{m}{2}\right) X\frac{1}{\sqrt{X^\dagger X}}\right).
 \ee
 It is known that the chiral mode then acquires a mass that is
 proportional to the parameter $m$ for small values of $m$.
 With these conventions, the lattice action describing
 a massive domain wall fermion depends on $4$ bare parameters:
 the Wilson mass parameter $M_0$, the temporal hopping parameter
 $\kappa_t$, the spatial Wilson parameter $r_s$ and the bare
 quark mass parameter $m$.

 In perturbation theory, it is convenient to study these
 matrices in Fourier space. After Fourier transformation of
 the matrix $X$:
 \be
  X_{xy} = \int^{\pi}_{-\pi}\frac{d^4p}{(2\pi)^4}\frac{d^4q}{(2\pi)^4}e^{i(qx -
  py)}\tilde{X}(q,p)\;,
  \ee
  the quantity $\tilde{X}(q,p)$ is expanded into
  power series in the bare coupling $g_0$:
  \be
  \tilde{X}(q,p) = \tilde{X}_0(p)(2\pi)^4 \delta_P(q - p)
  + \tilde{X}_1(q,p) + \tilde{X}_2(q,p) +
  O(g^3_0),
 \ee
 where the functions $\tilde{X}_i$ are of the order of $(g_0)^i$ respectively.
 Higher order contributions are not needed in an one-loop calculation.
 The explicit expressions are found to be:
 \ba
 \label{eq:Xs}
  \tilde{X}_0(p) &=& i\fslash{\tilde{p}} +
 \sum_\mu \kappa_\mu r_\mu(1- \cos
  p_\mu) + M_0,\label{eq:X_0}
  \nonumber \\
 \tilde{X}_1(q,p) &=& \sum_{A,\mu}\int \frac{d^4k}{(2\pi)^2} (2\pi)^4\delta_P(q-p-k)
 \times g_0 A_\mu^A(k)T^A V_{1\mu}\left(p+\frac{k}{2}\right),
 \nonumber \\
 \tilde{X}_2(q,p) &=& \sum_{A,B,\mu,\nu}\int^{\pi}_{-\pi}
 \frac{d^4k_1}{(2\pi)^4} \frac{d^4k_2}{(2\pi)^4}
 (2\pi)^4\delta_P(q-p-\sum  k_i)\nonumber\\
 &&\quad\times\frac{g^2}{2}A_\mu^A(k_1)A_\mu^B(k_2)T^AT^BV_{2\mu}\left(p+\frac{\sum
 k_i}{2}\right),
 \ea
 where $\fslash{\tilde{p}}=\sum_\mu\kappa_\mu\gamma_\mu\sin p_\mu$
 and the functions $V_{1\mu}$ and $V_{2\mu}$ are given by:
 \ba
 \label{eq:Vs}
 V_{1\mu}(p + \frac{k}{2}) &=&\kappa_\mu \left( i\gamma_\mu\cos \left(p +
 \frac{k}{2}\right)_\mu + r_\mu\sin \left(p  +
 \frac{k}{2}\right)_\mu \right)\;,
 \nonumber\\
 V_{2\mu}\left(p + \frac{\sum k_i}{2}\right) &=&\kappa_\mu
 \left( -i\gamma_\mu  \sin \left(p +
 \frac{\sum k_i}{2}\right)_\mu + r_\mu\cos \left(p + \frac{\sum
 k_i}{2}\right)_\mu \right)\;.
 \ea

 \section{Dispersion relation for the free overlap fermions}

 In this section, we briefly summarize the results for
 the free overlap fermions on anisotropic lattices.
 These results can be obtained from their counterparts for the isotropic lattices
 which can be found from the literature,
 see for example Ref.~\cite{neuberger98:massless_vector}.
 In the free case, we denote the overlap Dirac
 operator given in Eq.~(\ref{eq:overlapD_massive})
 by $D_0$. The inverse of $D_0$ in momentum space is
 found to be:
 \addtocounter{equation}{1}
 \begin{align}
  D_0^{-1}(p) &=
  \frac{1}{2}\frac{\left(1-\frac{m}{2}\right)X^\dagger_0(p)
  + \left(1+\frac{m}{2}\right)\omega(p)}
  {\left(1+\frac{m^2}{4}\right)\omega(p)
  +\left(1-\frac{m^2}{4}\right)b(p)},\label{eq:D_0inv}\tag{\theequation a}\\
 b(p) &= \sum_\mu \kappa_\mu r_\mu(1 - \cos p_\mu) + M_0,
 \label{eq:b_def}\tag{\theequation b}\\
  \omega(p) &= \sqrt{\tilde{p}^2 + \left(\sum_\mu
  \kappa_\mu r_\mu (1-\cos p_\mu) + M_0\right)^2}>0\;,
  \label{eq:omega_def}\tag{\theequation c}
 \end{align}
 where $\tilde{p}^2=\sum_\mu\kappa^2_\mu\sin^2p_\mu$.

 The dispersion relation for the physical particle
 (in this case, a free quark) is obtained by inspecting the
 time dependence of the propagator in coordinate space:
 \be
 G_0(\bp,t)=\int^\pi_{-\pi} {dp_0\over 2\pi} D^{-1}_0(p_0,\bp)e^{ip_0t}\;.
 \ee
 For large temporal separation $t$, the above propagator will
 behave like $e^{-tE_\bp}$ with $E_\bp$ being the energy of
 the particle with three-momentum $\bp$.
 Using the variable $z=e^{ip_0}$, the above propagator can also
 be expressed as:
 \be
 G_0(\bp,t)=\oint {dz\over 2\pi i}
 z^{t-1}D^{-1}(z,\bp)\;.
 \ee
 where the contour integral is along the unit circle
 counter-clockwise. If $D^{-1}(z,\bp)$ were an entire function
 of $z$, the above propagator receives contributions from
 the poles of $D^{-1}(z)$ within the unit circle.
 The case for the overlap is a bit more complicated due
 to the branch cut of $\omega(p)$. However, it can be shown
 that the leading contribution to the propagator $G_0(t,\bp)$
 at large $t$ comes from the residue of the pole closest
 to $z=1$ in the complex $z$ plane.
 Thus, setting the four-momentum:
 \be
 p_\mu=(E_\bp, \bp)\;,
 \ee
 and demanding that it is the pole of the propagator $D_0^{-1}(p)$,
 the above equations yield the relation:
 \be
 -\left(1+\frac{m^2}{4}\right)^2\tilde{p}^2=m^2b^2
 \;.
 \ee
 We are only interested in the dispersion relation when
 the lattice three-momentum $|\bp|\ll 1$. We thus obtain:
 \be
 E_\bp=E_0+{\bp^2\over 2M_{kin}}+ O(\bp^4)\;,
 \ee
 where $E_0\equiv m_Q$ will be identified as the pole
 mass of the quark and $M_{kin}$ is
 the so-called kinetic mass of the quark.
 After some calculations, the equation satisfied
 by $E_0=m_Q$ is found to be:
 \be
 \label{eq:mQ}
 A\cosh^2E_0+B\cosh E_0+C=0\;,
 \ee
 where the coefficients are given by:
 \begin{equation}
 \label{eq:ABC}
 \begin{split}
  A&=\kappa_t^2\left(1+\frac{m^2}{4}\right)^2 - \kappa_t^2 m^2\;,\\
 B&=2 \kappa_t m^2 \left( \kappa_t+M_0\right)\;, \\
 C&=-\kappa_t^2\left(1+\frac{m^2}{4}\right)^2-m^2\left(
 \kappa_t+M_0\right)^2\;.
  \end{split}
  \end{equation}
 It is seen that $E_0\equiv m_Q$ is independent of the parameter
 $r_s$. It is also easy to verify that for small values of $m$, the
 solution $E_0=m_Q$ obtained from Eq.~(\ref{eq:mQ}) is proportional to $m$,
 as expected. We can also find out the kinetic mass term with the result:
 \be
 {1\over 2M_{kin}}=\frac{d E_\bp}{d\bp^2}\Bigg|_{\bp=0}=
 \frac{\left(1+\frac{m^2}{4}\right)^2+
 m^2 r_s \left(M_0+ \kappa_t- \kappa_t\cosh E_0\right)}
 {(2A\cosh E_0+B)\sinh E_0}
 \ee
 In order to have the usual energy-momentum
 dispersion relation for the quark, one imposes the condition:
 $M_{kin}=E_0\equiv m_Q$ which yields the relationship between
 $\chi$ and $\kappa_t$ as
 (restoring the lattice spacing explicitly):
 \be
 \label{eq:chi}
 \chi^2=\frac{(A\cosh E_0+B/2)\left(\sinh E_0/E_0\right)}
 {\left(1+\frac{m^2}{4}\right)^2+
 m^2 r_s \left(M_0+ \kappa_t- \kappa_t\cosh E_0\right)}
 \;,
 \ee
 where the value of $E_0=m_Q$ is to be obtained from
 Eq.~(\ref{eq:mQ}).
 To summarize, for a given set of parameters $\chi$, $M_0$, $m$ and
 $r_s$, one has to solve both Eq.~(\ref{eq:mQ}) and
 Eq.~(\ref{eq:chi}) to obtain the values of $\kappa_t$ and $m_Q$.

 \begin{figure}[htb]
 \begin{center}
 \includegraphics[width=12.0cm,angle=0]{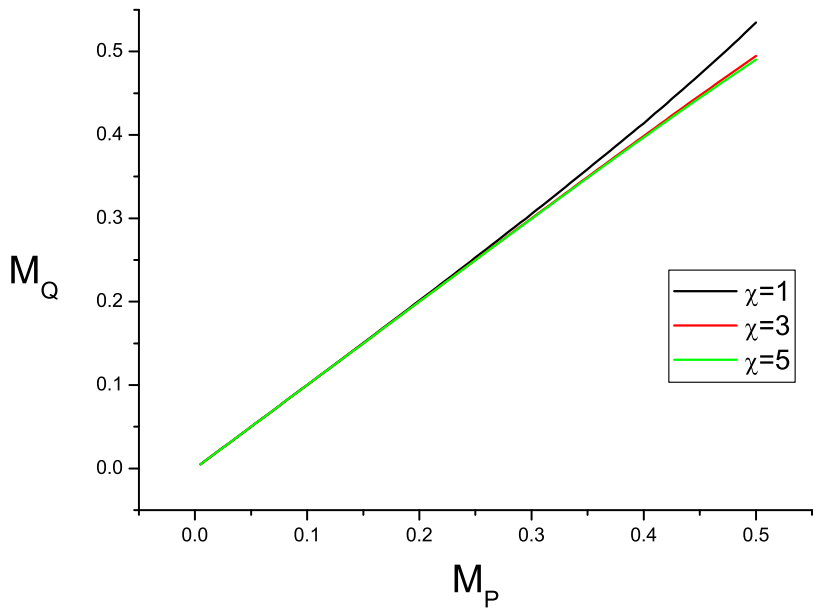}
 \end{center}
 \caption{The pole mass of the free domain wall quark, $m_Q$, measured
 in $1/a_s$ unit as a function of the propagator bare mass parameter $m_P=m|M_0|$
 for three values of the anisotropy parameter $\chi$. Three curves
 correspond to $\chi=1$, $3$ and $5$, respectively.
 We have set $M_0=-0.5$ and $r_s=1$ in this plot.}
 \label{fig:mQvsmP}
 \end{figure}
 In Fig.~\ref{fig:mQvsmP}, we have shown the pole mass of the
 quark as a function of the propagator mass parameter: $m_P=m|M_0|$ for
 a given set of other parameters. Three curves in the plot correspond to $\chi=1,3,5$
 respectively. The value of $\kappa_t$ is obtained from Eq.~(\ref{eq:chi}).
 It is seen that the pole mass of the quark increases with
 $m_P$ linearly for small values of $m_P$. For larger values
 of bare quark mass, some non-linearity sets in.

 \begin{figure}[htb]
 \begin{center}
 \includegraphics[width=12.0cm,angle=0]{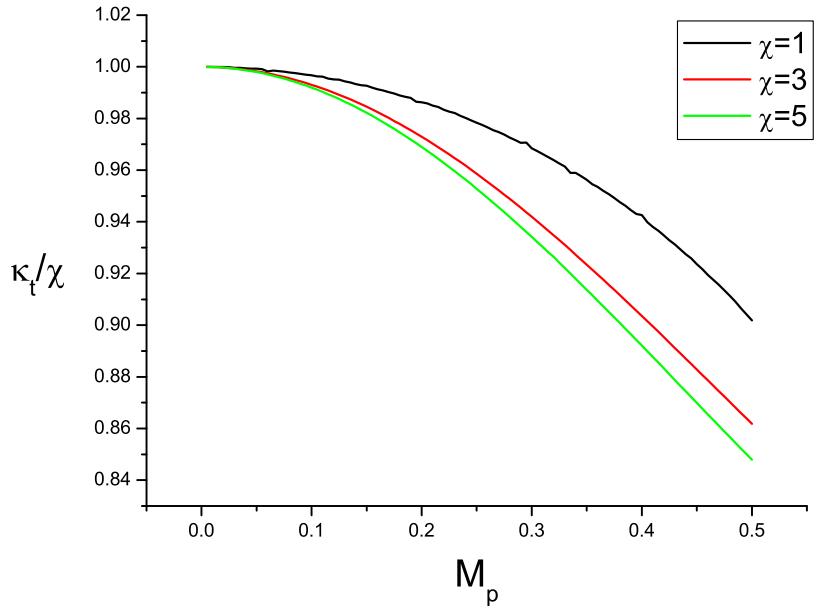}
 \end{center}
 \caption{With the same set of parameters as Fig.~\ref{fig:mQvsmP},
 the value of $\kappa_t/\chi$ is shown as
 a function of the propagator mass parameter $m_P$ for $\chi=1$, $3$ and $5$.
 It is seen that in the massless limit,
 the expected relation: $\chi=\kappa_t$ is recovered for all $\chi$.}
 \label{fig:KtvsmQ}
 \end{figure}
 In Fig.~\ref{fig:KtvsmQ}, for a given value of anisotropy $\chi$,
 we have plotted the appropriate value of $\kappa_t$
 as a function of the propagator mass parameter $m_P$.
 Since $m_P$ is very close to the pole mass $m_Q$ as
 Fig.~\ref{fig:mQvsmP} indicates,
 this figure can also be viewed as the
 dependence of $\kappa_t$ on $m_Q$ for a given anisotropy $\chi$.
 It is seen that, in the massless limit, i.e. $m_Q\rightarrow 0$,
 one recovers the naive relation: $\chi=\kappa_t$.
 However for massive quarks, this relation is distorted
 with increasing values of the quark mass. The deviation from
 its massless limit can be as large as $15$\% for
 values of $m_P\simeq 0.5$.
 In quenched Monte Carlo simulations, the
 anisotropy $\chi$ is fixed by the pure gauge sector.
 Therefore, this figure can be utilized  to tune the
 hopping parameter $\kappa_t$
 accordingly for a given value of the quark mass and the anisotropy.
 In a non-perturbative Monte Carlo simulation, this tuning process can be
 performed by demanding physical particles, e.g. the pions, have the
 correct dispersion relation for small three-momenta.
 We would like to emphasize that
 this tuning process is crucial in the simulation of massive quarks
 since, without it, the hadrons will not have the correct continuum limit.

 In both Fig.~\ref{fig:mQvsmP} and Fig.~\ref{fig:KtvsmQ} we have
 chosen the Wilson parameters $r_s$ to be unity.
 In principle, other values are also permitted as long as the doublers
 are well separated from the physical modes.
 It is seen from Eq.~(\ref{eq:b_def}) that the separation
 between the doublers and the normal physical quark
 is characterized by the Wilson parameter $r_s$.
 Therefore, to decouple the doublers it is better not to
 take too small values for $r_s$.
 In the remaining part of this paper, we will
 take the conventional choice: $r_t=r_s=1$.

 \section{The wave function and quark mass renormalization to one-loop}

 In this section, we will compute the quark self-energy
 to one-loop using bare perturbation theory. Two diagrams
 contribute at this level: the tadpole diagram and the
 half-circle diagram. We will first list the
 necessary Feynman rules and the calculation of
 these two diagrams will be dealt with afterwards.

 \subsection{Feynman rules}

 First of all, one needs
 the free propagator of the lattice gauge
 fields~\cite{shigemitsu00:aniso_oneloop}.
 After performing Fourier transformation of the gauge fields,
 the quadratic part of the gauge action has the
 standard form in momentum space:
  \ba
 S^{(0)}_g[A_\mu] = \frac{1}{2} \, \sum_{\mu\nu}
 \int_{-\pi}^\pi \frac{d^4 l}{(2 \pi)^4} \left( \bar{A}_\mu(l)
 \,M_{\mu\nu}(l) \, \bar{A}_\nu(-l) \right) ,
 \ea
 where
 \ba
 M_{00} &=& \chi \, \left[ \frac{\chi^2}{\alpha_g} \hat{l}_0^2 \; + \;
 \sum_j
 \hat{l}_j^2 \, q_{0j} \right]   \\
 M_{jj} &=& \frac{1}{\chi} \, \left[ \frac{1}{\alpha_g} \hat{l}_j^2 \; +
 \; \chi^2 \, \hat{l}_0^2 \, q_{0j} \; + \; \sum_{j'\neq j}\hat{l}_{j'}^2 \,
 q_{j'j}
 \right]   \\
 M_{i \neq j} &=& \frac{1}{\chi} \, \left[ \frac{1}{\alpha_g}
 \hat{l}_i \hat{j}_j \; - \;  \hat{l}_i \hat{l}_j \, q_{ij} \right]   \\
 M_{0 j} &=& M_{j 0}  = \chi \, \left[ \frac{1}{\alpha_g}
 \hat{l}_0 \hat{l}_j \; - \;  \hat{l}_0 \hat{l}_j \, q_{0j} \right]
 \ea
 with lattice momentum defined as:
 $\hat{l}_\mu \equiv 2\,\sin(\frac{l_\mu}{2})$.
 The quantities $q_{\mu\nu}$ appearing in the above equations
 are given by:
 \be
  q_{0j} = 1 \; + \; \frac{1}{12} \, \hat{l}^2_j\;,
  \;\;
  q_{ij} = 1 \; + \; \frac{1}{12} \, (\hat{l}^2_i + \hat{l}^2_j)
 \qquad i \neq j \;.
 \ee
 Using these notations, the free gluon propagator is expressed as:
 \ba
 \label{eq:freeGpropagator}
 D_{\mu\nu}(l) =
 M^{-1}_{\mu\nu}
 = \frac{1}{(\hat{l}^2)^2} \left[ \alpha_g \hat{l}_\mu
 \hat{l}_\nu \chi +
 \frac{f^{\mu\nu}(\hat{l}_\rho,q_{\rho\sigma},\chi)}
 {f_D(\hat{l}_\rho,q_{\rho\sigma},\chi)} \right] ,
 \ea
 The explicit expressions for $f^{\mu\nu}$ and  $f_D$ maybe found
 in the literature~\cite{shigemitsu00:aniso_oneloop}.
 For simplicity,  in the following calculation we
 choose the gauge in which $\alpha_g=1$.

 The vertex functions for the interaction between the quark
 and the gluon fields can also be obtained with the help of
 Eq.~(\ref{eq:Xs}) and Eq.~(\ref{eq:Vs}). The explicit expressions
 are given in Eq.~(\ref{eq:vertex3}) and Eq.~(\ref{eq:vertex4}) in the appendix.

 \subsection{The half-circle diagram}

 For the half-circle diagram, the contribution to
 the quark self-energy can be written as:
 \ba
 \label{eq:half}
 \Sigma^{\rm half-circle}(p)&=&g^2_0C_F \int
 \frac{d^4 k}{(2 \pi)^4}\sum_\mu
 \frac{\left(1-\frac{m}{2}\right)^2}{\left\{\omega(k)
  +\omega(p)\right\}^2}
 \nonumber \\
  &\times&\left\{  \frac{1}{2}\frac{\sigma^{(1)}_\mu(p,k)}
  {\left(1+\frac{m^2}{4}\right)\omega(k)
  +\left(1-\frac{m^2}{4}\right)b(k)}\right\}D_{\mu\mu}(p-k)
  \;,
 \ea
 where
 \ba
 \sigma^{(1)}_\mu(p,k)&=&\left\{V_{1\mu}\left(\frac{p+k}{2}\right) -
 \frac{X_0(p)}{\omega(p)}V_{1\mu}^{\dagger}\left(\frac{p+k}{2}\right)\frac{X_0(k)}
 {\omega(k)}
 \right\}\nonumber\\
 &\times& \left\{  \left(1-\frac{m}{2}\right)X^\dagger_0(k)
  + \left(1+\frac{m}{2}\right)\omega(k)\right\}
 \nonumber \\
 &\times&
  \left\{V_{1\mu}\left(\frac{p+k}{2}\right) -
 \frac{X_0(k)}{\omega(k)}V_{1\mu}^{\dagger}\left(\frac{p+k}{2}\right)\frac{X_0(p)}
 {\omega(p)}
 \right\}\;,
 \ea
 and $D_{\mu\mu}$ is the gluon propagator given in
 Eq.~(\ref{eq:freeGpropagator}).

 We will be interested in the small $p$ behavior of the self-energy.
 Therefore, one can expand the integrand in Eq.~(\ref{eq:half}) around $p=0$.
 Note that $\omega(p)$ is an even function of $p$. Therefore,
 when expanded around $p=0$, one has: $\omega(p)=\omega(0)+O(p^2)$.
 However, both $D_{\mu\mu}(p-k)$ and $\sigma^{(1)}_\mu(p,k)$
 contains terms that are linear in $p$.
 When expanding the function $D_{\mu\mu}(p-k)$,
 the leading term is an even function of $k$, the term
 proportional to $p$ is an odd function of $k$.
 Therefore, when multiplied with expansion of the function
 $\sigma^{(1)}_\mu(p,k)$, only the terms that are
 even functions of $k$ will give non-zero contributions after the integration
 over $k$ is performed.
 For example, if we take $D_{\mu\mu}(k)$, then
 it suffice to keep only even functions of $k$ in the
 expansion of $\sigma^{(1)}_\mu(p,k)$:
 \ba
 \sigma^{(1)}_\mu(0,k)&=&\left\{V_{1\mu}\left(\frac{k}{2}\right) +
 V_{1\mu}^{\dagger}\left(\frac{k}{2}\right)\frac{X_0(k)}
 {\omega(k)}
 \right\}\nonumber\\
 &\times& \left\{  \left(1-\frac{m}{2}\right)X^\dagger_0(k)
  + \left(1+\frac{m}{2}\right)\omega(k)\right\}
 \nonumber \\
 &\times&
  \left\{V_{1\mu}\left(\frac{k}{2}\right) +
 \frac{X_0(k)}{\omega(k)}V_{1\mu}^{\dagger}\left(\frac{k}{2}\right)
 \right\}\;,
 \nonumber \\
 &\sim&\left\{\left(1-\frac{m}{2}\right)\omega+\left(1+\frac{m}{2}\right)b\right\}
 \nonumber \\
 &\times&
 \left(V_{1\mu}^{\dagger}V_{1\mu}
 +V_{1\mu}V_{1\mu}^{\dagger}+\frac{2}{\omega}V_{1\mu}^{\dagger}X_0V_{1\mu}^{\dagger}\right)
 \;.
 \ea
 Here the symbol ``$\sim$" means that the odd functions of $k$ are dropped
 from the expression. Similarly, if the terms linear in $p$ from
 $D_{\mu\mu}(p-k)$ are taken, which is an odd function of $k$, then
 only terms that are odd functions of $k$ in the expansion
 of $\sigma^{(1)}_\mu(p,k)$ will contribute
 to the integral.

 Then, we calculate the quantity $d\sigma^{(1)}_\nu(p,k)/dp_\mu$.
 First, we notice that
 \ba
 \left.\frac{dV_{1\nu}\left(\frac{p+k}{2}\right)}{dp_\mu}\right|_{p=0}=
 -\frac{i}{2}\gamma_\mu V_{1\mu}\left(\frac{k}{2}\right)\delta_{\mu\nu}
 \;,\quad
 \left.\frac{dX_0(p)}{dp_\mu}\right|_{p=0}=i\kappa_\mu\gamma_\mu \;.
 \ea
 We define:
 $d\sigma^{(1)}_\nu(p,k)/dp_\mu\equiv
 i\kappa_\mu\gamma_\mu\sigma^{(1)\nu}_\mu$ and
 the explicit calculation shows:
 \ba
 \left.\frac{d\sigma^{(1)}_\nu(p,k)}{dp_\mu}\right|_{p=0}&=&
 \left\{-\frac{i}{2}\gamma_\mu V_{1\mu}\left(\frac{k}{2}\right)\delta_{\mu\nu} +
 i\gamma_\mu\frac{\kappa_\mu}{M_0}V_{1\nu}^{\dagger}\left(\frac{k}{2}\right)\frac{X_0(k)}
 {\omega(k)}\right.
 \nonumber \\
 &&+\left.\frac{i}{2}\gamma_\mu V_{1\mu}^{\dagger}\left(\frac{k}{2}\right)\frac{X_0(k)}
 {\omega(k)}\delta_{\mu\nu}
 \right\}\nonumber\\
 &\times& \left\{  \left(1-\frac{m}{2}\right)X^\dagger_0(k)
  + \left(1+\frac{m}{2}\right)\omega(k)\right\}
 \nonumber \\
 &\times&
  \left\{V_{1\nu}\left(\frac{k}{2}\right) +
 \frac{X_0(k)}{\omega(k)}V_{1\nu}^{\dagger}\left(\frac{k}{2}\right)
 \right\}
 \nonumber \\
 &+&\left\{V_{1\nu}\left(\frac{k}{2}\right) +
 V_{1\nu}^{\dagger}\left(\frac{k}{2}\right)\frac{X_0(k)}
 {\omega(k)}
 \right\}\nonumber\\
 &\times& \left\{  \left(1-\frac{m}{2}\right)X^\dagger_0(k)
  + \left(1+\frac{m}{2}\right)\omega(k)\right\}
 \nonumber \\
 &\times&
  \left\{-\frac{i}{2}\gamma_\mu V_{1\mu}\left(\frac{k}{2}\right)\delta_{\mu\nu} +
 \frac{X_0(k)}{\omega(k)}V_{1\nu}^{\dagger}\left(\frac{k}{2}\right)
 i\gamma_\mu\frac{\kappa_\mu}{M_0}\right.
 \nonumber \\
 &&\left.+\frac{X_0(k)}{\omega(k)}V_{1\mu}^{\dagger}\left(\frac{k}{2}\right)
 \frac{i}{2}\gamma_\mu\delta_{\mu\nu}
 \right\}\;.
 \ea
 As we clarified before, we only need to keep odd functions
 of $k$ in this quantity. Thus we get:
 \ba
 \left.\frac{d\sigma^{(1)}_\nu(p,k)}{dp_\mu}\right|_{p=0}
 &\sim&\left(1-\frac{m}{2}\right)
 \left\{-\frac{i}{2}\{\gamma_\mu,V_{1\mu}X_0^\dagger V_{1\mu}\}\delta_{\mu\nu}
 +\frac{i}{2}\{\gamma_\mu,V_{1\mu}^\dagger X_0V_{1\mu}^\dagger\}\delta_{\mu\nu}\right.
 \nonumber \\
 &&\left.+i\gamma_\mu\frac{\kappa_\mu}{M_0}\omega\left(V_{1\nu}^\dagger
 V_{1\nu}+V_{1\nu}V_{1\nu}^\dagger\right)
 +i\frac{\kappa_\mu}{M_0}\{\gamma_\mu,V_{1\nu}^\dagger
 X_0V_{1\nu}^\dagger\}\right\}
 \nonumber \\
 &+&\left(1+\frac{m}{2}\right)\Biggl\{-i\gamma_\mu\omega
 \left(V_{1\mu}V_{1\mu}+V^{\dagger}_{1\mu}V^{\dagger}_{1\mu}\right)\delta_{\mu\nu}
 \nonumber \\
 &&+i\frac{b}{\omega}\{\gamma_\mu,V_{1\mu}^{\dagger}X_0V_{1\mu}^{\dagger}\}\delta_{\mu\nu}
 +i\gamma_\mu\frac{\kappa_\mu}{M_0}b
 \left(V_{1\nu}^{\dagger}V_{1\nu}+V_{1\nu}V_{1\nu}^{\dagger}\right)
 \nonumber \\
 &&-i\gamma_\mu\frac{\kappa_\mu}{M_0}2\omega
 V_{1\nu}^{\dagger}V_{1\nu}^{\dagger}+i\frac{\kappa_\mu}{M_0}\frac{2b}{\omega}
 \{\gamma_\mu,V_{1\nu}^{\dagger}X_0V_{1\nu}^{\dagger}\}\Biggr\}
 \;.
 \ea
 Here the symbol ``$\sim$" means that the even functions of $k$ are dropped
 from the expression. Using the expressions for various quantities,
 we have:
 \ba
 &&V_{1\nu}^\dagger
 V_{1\nu}=V_{1\nu}V_{1\nu}^\dagger=\kappa_\nu^2
 \nonumber \\
 &&V_{1\nu}V_{1\nu}\sim-\kappa_\nu^2\cos k_\nu\quad
 V_{1\nu}^{\dagger}V_{1\nu}^{\dagger}\sim-\kappa_\nu^2\cos k_\nu
 \nonumber \\
 &&V_{1\nu}^{\dagger}X_0V_{1\nu}^{\dagger}\sim\kappa_\nu^2
 \left(\kappa_\nu\sin^2k_\nu-b\cos k_\nu\right)
 \nonumber \\
 &&V_{1\nu}X_0^{\dagger}V_{1\nu}\sim\kappa_\nu^2
 \left(\kappa_\nu\sin^2k_\nu-b\cos k_\nu\right)
 \ea
 Finally, we calculate the term $dD_{\nu\nu}(p-k)/dp_\mu|_{p=0}$:
 \ba
 \left.\frac{dD_{\nu\nu}(p-k)}{dp_\mu}\right|_{p=0}&=&\hat{k}_\mu\Biggl\{\frac{4\left(\chi^2\right)^{\delta_{\mu 0}}}
 {\left(\chi^2\hat{k_0}^2+\sum_j\hat{k}_j^2\right)^3}
 \left[\chi\hat{k_\nu}^2+\frac{f^{\nu\nu}}{f_D}\right]
 \nonumber \\
 &+&\frac{1}{\left(\chi^2\hat{k_0}^2+\sum_j\hat{k}_j^2\right)^2}
 \left[-2\chi\delta_{\mu\nu}-\frac{f^{\nu}_{\mu}}{f_D}+\frac{f^D_\mu
 f^{\nu\nu}}{f_D^2}\right]\Biggr\}
 \ea
 where we have used the following notations:
 \be
 \label{eq:fmunu}
 f^\nu_\mu(k) = -\left.\frac{\partial f^{\nu\nu}(p-k)}{\hat{k}_\mu \partial p_\mu}\right|_{p=0}, \quad
 f^D_\mu(k) =-\left.\frac{\partial f_D (p-k)}{\hat{k}_\mu \partial
 p_\mu}\right|_{p=0}\;.
 \ee
 As mentioned above, here the even functions of $k$ in $\sigma^{(1)}_\nu(0,k)$ will
 vanish and the odd ones will survive.
 These terms are of the following form:
 \ba
 \hat{k}_\mu\sigma^{(1)}_\nu(0,k)&\sim& i\kappa_\mu\gamma_\mu\hat{k}_\mu\sin k_\mu
 \Biggl\{
 2\left(1+\frac{m}{2}\right)\kappa_\mu\omega\delta_{\mu\nu}
 \nonumber \\
 &&+2\left(1+\frac{m}{2}\right)\left(\kappa_\nu^2\sin^2\frac{k_\nu}{2}
 +\kappa^2_\nu\cos^2\frac{k_\nu}{2}\left(2\delta_{\mu\nu}-1\right)\right)
 \nonumber \\
 &&+\frac{2b}{\omega}\left(1+\frac{m}{2}\right)
 \left(-b\kappa_\mu\delta_{\mu\nu}+\kappa_\nu^2-2\kappa_\mu^2\cos^2
 \frac{k_\mu}{2}\delta_{\mu\nu}\right)\Biggr\}
 \ea
 Collecting all the relevant terms, we
 may write the contribution from the half-circle diagram as:
 \ba
 \label{eq:halfcircle}
 \Sigma^{\rm half-circle}(p)& = &g^2\left(i \sum_\mu\gamma_\mu\tilde{p}_\mu
 I_\mu^{(1)}+M^{(1)} \right)\;.
 \ea
 Introducing  the following notations:
 \ba
 \left.\frac{d\sigma^{(1)}_\nu(p,k)}{dp_\mu}\right|_{p=0}=i\kappa_\mu\gamma_\mu
 \overline{\sigma}^{(1)\nu}_\mu(k)\quad \hat{k}_\mu\sigma^{(1)}_\nu(0,k)=i\kappa_\mu\gamma_\mu
 \widetilde{\sigma}^{(1)\nu}_\mu(k)\;,
 \ea
 the quantity $I^{(1)}$ and $M^{(1)}$ are expressed
 in terms of the following integrals:
 \ba
 I^{(1)}_\mu&=&C_F \int
 \frac{d^4 k}{(2 \pi)^4}\sum_\nu
 \frac{\left(1-\frac{m}{2}\right)^2}{\left\{\omega
  -M_0\right\}^2}
 \left\{  \frac{1}{2}\frac{1}
  {\left(1+\frac{m^2}{4}\right)\omega
  +\left(1-\frac{m^2}{4}\right)b}\right\}
  \nonumber \\
  &\times&
  \left(\overline{\sigma}^{(1)\nu}_\mu(k)S_g^{\nu\nu}(k)+
  \widetilde{\sigma}^{(1)\nu}_\mu(k)
  \frac{dD_{\nu\nu}(p-k)}{\hat{k}_\mu dp_\mu}|_{p=0}\right)
  \;,
  \\
 M^{(1)}&=&C_F \int
 \frac{d^4 k}{(2 \pi)^4}\sum_\mu
 \frac{\left(1-\frac{m}{2}\right)^2}{\left\{\omega
  -M_0\right\}^2}
 \nonumber \\
  &\times&\left\{  \frac{1}{2}\frac{\sigma^{(1)}_\mu(0,k)}
  {\left(1+\frac{m^2}{4}\right)\omega
  +\left(1-\frac{m^2}{4}\right)b}\right\}D_{\mu\mu}(k)
  \;.
 \ea
 where all integrations are taken in the first Brillouin zone.

 \subsection{The tadpole diagram}

 Using the Feynman rules given in the previous section,
 the contribution from the tadpole diagram can be written as:
 \ba
 \label{eq:tadpole}
 \Sigma^{\rm tadpole}(p) &=& \frac{1}{2} g^2 C_F
 \int_{-\pi}^\pi \frac{d^4 k}{(2 \pi)^4}\sum_\mu
 \left(1-\frac{m}{2}\right)\sigma^{(2)}_\mu(p,k)D_{\mu\mu}(k)
 \;,
 \ea
 where the function $\sigma^{(2)}_\mu(p,k)$ reads:
 \ba
 \sigma^{(2)}_\mu(p,k)&=&-\frac{1}{2\omega(p)}\left\{ V_{2\mu}(p) -
  \frac{X_0(p)}{\omega(p)}V_{2\mu}^\dagger(p)\frac{X_0(p)}{\omega(p)}\right\}
 +\frac{1}{\omega(p)\left\{\omega(p) + \omega(p+k)\right\}^2}\nonumber \\
 &&\times\left\{ X_0(p)V^\dagger_{1\mu}\left(p+
  \frac{k}{2}\right)V_{1\mu}\left(p + \frac{k}{2}\right) + V_{1\mu}\left(p+
  \frac{k}{2}\right)X_0^\dagger(p+k) V_{1\mu}\left(p + \frac{k}{2}\right)\right.\nonumber \\
 &&  \hspace{0.5cm}+  V_{1\mu}\left(p+\frac{k}{2}\right) V_{1\mu}^\dagger
 \left(p + \frac{k}{2}\right)X_0(p)\nonumber \\
 &&  \hspace{0.5cm}-
  \left.\frac{2\omega(p)+\omega(p+k)}{\omega(p)^2\omega(p+k)}X_0(p)
  V_{1\mu}^\dagger\left(p+\frac{k}{2}\right)X_0(p+k)V_{1\mu}^\dagger
  \left(p + \frac{k}{2}\right)X_0(p)\right\}\nonumber.\\
 \ea
 For small lattice momenta $p$, the function
 $\sigma^{(2)}_\mu(p,k)$ is expanded. It contains
 a term at vanishing momentum:
 \ba
 \sigma^{(2)}_\mu(0,k)=-\frac{1}{\{\omega-M_0\}^2}\left( V^\dagger_{1\mu}V_{1\mu} +
 V_{1\mu} V_{1\mu}^\dagger + \frac{2}{\omega}V_{1\mu}^\dagger X_0
 V_{1\mu}^\dagger\right)\;,
 \ea
 and a term that is linear in $p$.
 To evaluate this term,
 we need $d\sigma^{(2)}_\nu(p,k)/dp_\mu\equiv i\kappa_\mu\gamma_\mu\sigma^{(2)\nu}_\mu$:
 \ba
 \left.\frac{d\sigma^{(2)}_\nu(p,k)}{dp_\mu}\right|_{p=0}
 &=&-\frac{i\kappa_\mu\gamma_\mu}{M_0}
 \left(\delta_{\mu\nu}+\frac{\kappa_\nu}{M_0}\right)-i\kappa_\mu\gamma_\mu
 \frac{4\kappa_\mu^2\cos
 k_\mu+4\kappa_\mu b}{\{\omega-M_0\}^2M_0\omega^2}\sin^2k_\mu
 \nonumber \\
 &\times&\left\{-b\kappa_\mu\delta_{\mu\nu}+
 \kappa_\nu^2
 -2\kappa_\mu^2\cos^2\frac{k_\mu}{2}\delta_{\mu\nu}\right\}
 \nonumber \\
 &-&\frac{1}{M_0\{\omega(k)-M_0\}^2}\Biggl[i\kappa_\mu\gamma_\mu
 \left(V_{1\nu}^{\dagger}V_{1\nu}+V_{1\nu}V_{1\nu}^{\dagger}\right)
 \nonumber \\
 &-&i\{\gamma_\mu,V_{1\mu}X_0^{\dagger}V_{1\mu}\}\delta_{\mu\nu}
 +\left.V_{1\nu}\frac{dX_0^{\dagger}(p+k)}{dp_\mu}\right|_{p=0}V_{1\nu}
 \nonumber \\
 &-&i\kappa_\mu\gamma_\mu\frac{2\kappa_\mu^2\cos
 k_\mu+2\kappa_\mu b}{\omega^3}\sin^2k_\mu
 \nonumber \\
 &\times&M_0\left\{-b\kappa_\mu\delta_{\mu\nu}+
 \kappa_\nu^2
 -2\kappa_\mu^2\cos^2\frac{k_\mu}{2}\delta_{\mu\nu}\right\}
 \nonumber \\
 &-&\left(\frac{1}{M_0}-\frac{2}{\omega}\right)\Biggl(i\kappa_\mu
 \{\gamma_\mu,V_{1\nu}^{\dagger}X_0V_{1\nu}^{\dagger}\}
 +iM_0\{\gamma_\mu,V_{1\mu}^{\dagger}X_0V_{1\mu}^{\dagger}\}\delta_{\mu\nu}
 \nonumber \\
 &+&M_0V_{1\nu}^{\dagger}\frac{dX_0(p+k)}{dp_\mu}|_{p=0}V_{1\nu}^{\dagger}\Biggr)\Biggr]
 \;.
 \ea
 Note that we have:
 \ba
 &&\left.V_{1\nu}\frac{dX_0^{\dagger}(p+k)}{dp_\mu}\right|_{p=0}V_{1\nu}
 \sim-i\kappa_\mu\gamma_\mu\cos
 k_\mu\kappa_\nu^2+i\kappa_\mu^3\gamma_\mu(1+\cos k_\mu)\delta_{\mu\nu}
 \;,
 \nonumber \\
 &&\left.V_{1\nu}^{\dagger}\frac{dX_0(p+k)}{dp_\mu}\right|_{p=0}V_{1\nu}^{\dagger}
 =-V_{1\nu}\frac{dX_0^{\dagger}(p+k)}{dp_\mu}|_{p=0}V_{1\nu}
 \;.
 \ea
 Therefore, if we introduce the following denotation
 \ba
 \left.\frac{d\sigma^{(2)}_\nu(p,k)}{dp_\mu}\right|_{p=0}=i\kappa_\mu\gamma_\mu
 \overline{\sigma}^{(2)\nu}_\mu(k)
 \;,
 \ea
 we can parameterize the contribution from the tadpole diagram as:
 \ba
 \Sigma^{\rm tadpole}(p)=g^2\left(i
 \sum_\mu\gamma_\mu\tilde{p}_\mu
 I_\mu^{(2)}+M^{(2)} \right)\;,
 \ea
 where the quantities $I^{(2)}$ and $M^{(2)}$ are given by:
 \ba
 I_\mu^{(2)}&=& \frac{1}{2}C_F
 \int_{-\pi}^\pi \frac{d^4 k}{(2 \pi)^4}\sum_\nu
 \left(1-\frac{m}{2}\right)\overline{\sigma}^{(2)\nu}_\mu(k)
 D_{\nu\nu}(k)
 \;,
 \\
 M^{(2)}&=&\frac{1}{2}C_F
 \int_{-\pi}^\pi \frac{d^4 k}{(2 \pi)^4}\sum_\mu
 \left(1-\frac{m}{2}\right)\sigma^{(2)}_\mu(0,k)D_{\mu\mu}(k)
 \;.
 \ea

 Adding the contributions from the half circle and the
 tadpole diagrams together, one can verify that
 $M^{(1)}+M^{(2)}$ is proportional to the quark mass:
 \ba
 M^{(1)}+M^{(2)}&=&-C_F \int_{-\pi}^\pi
 \frac{d^4 k}{(2 \pi)^4}\sum_\mu
 \frac{m\left(1-\frac{m}{2}\right)}{\left\{\omega
  -M_0\right\}^2}
 \nonumber \\
  &\times&\left\{  \frac{1}{2}\frac{\omega
  \left(V_{1\mu}^{\dagger}V_{1\mu}+V_{1\mu}V_{1\mu}^{\dagger}
  +\frac{2}{\omega}V_{1\mu}^{\dagger}X_0V_{1\mu}^{\dagger}\right)}
  {\left(1+\frac{m^2}{4}\right)\omega
  +\left(1-\frac{m^2}{4}\right)b}\right\}D_{\mu\mu}(k)
  \;.
 \ea
 This is the expected result since it is known that
 quark mass is only multiplicative renormalized in QCD due
 to chiral symmetry. In what follows, we will define:
 \be
 M^{(3)}=\frac{M^{(1)}+M^{(2)}}{m_P}\;,
 \ee
 which is a finite quantity in the chiral limit.

 \subsection{The massless case}

 If the quark mass is exactly zero, then the one-loop contribution
 discussed above contains infra-red divergences.
 However, the same infra-red divergence
 also exists in the continuum. Therefore, one can subtract an appropriate
 infra-red divergent part from the one-loop contribution:
 \ba
 \label{eq:Ifinite}
 I_{\rm finite,\mu}^{(1)}&=&I_\mu^{(1)}(m=0)-\frac{C_F}{M_0}
 \int_{-\pi}^{\pi} \frac{d^4 k}{(2\pi)^4}
 \frac{4\chi(\chi^2)^{\delta_{\mu0}}k^2_\mu\theta
 (\pi^2-(\chi^2k^2_0+\sum_j k^2_j))}
 {(\chi^2k^2_0+\sum_j k^2_j)^3}
 \;
 \\
 \label{eq:Ilog}
 I_{\rm log,\mu}^{(1)}&=&\frac{C_F}{16\pi^2M_0}
 \Biggl( \ln (\pi^2) + 1 -\ln(\tilde{p}^2) \Biggr)
 \ea
 \ba
 \label{eq:Mfinite}
 M_{\rm finite}^{(3)}&=&M^{(3)}(m=0)-\frac{C_F}{M_0}
 \int_{-\pi}^{\pi} \frac{d^4 k}{(2\pi)^4}
 \frac{4\chi\theta
 (\pi^2-(\chi^2k^2_0+\sum_j k^2_j))}
 {(\chi^2k^2_0+\sum_j k^2_j)^2}
 \;
 \\
 M_{\rm log}^{(3)}&=&\frac{C_F}{4\pi^2M_0}
 \Biggl( \ln (\pi^2) + 1 -\ln(\tilde{p}^2)\Biggr)
 \ea
 The subtracted contributions listed above are all infra-red finite.
 Therefore they can be evaluated numerically once the bare parameters are fixed.
 The infra-red divergent part can be obtained analytically and depends on
 the external momentum (or a scale parameter when external momentum
 is vanishing) as usual.

 \subsection{Numerical results}

 \begin{figure}[htb]
 \begin{center}
 \includegraphics[width=12.0cm,angle=0]{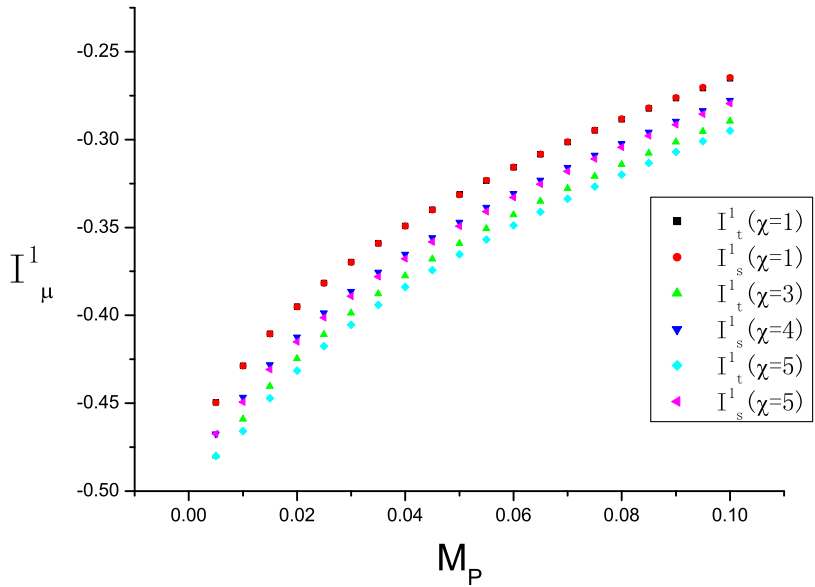}
 \end{center}
 \caption{The values of $I^{(1)}_\mu(m)$ as a function of $m_P$ are
 shown for $\chi=1$, $3$ and $5$.  Here $I^{(1)}_t$ corresponds
 to $\mu=0$ and $I^{(1)}_s$ corresponds to $\mu=1,2,3$.
 Other bare parameters are:
 $\kappa_s=1$,  $M_0=-0.5$.
 \label{fig:I1}}
 \end{figure}
 \begin{figure}[htb]
 \begin{center}
 \includegraphics[width=12.0cm,angle=0]{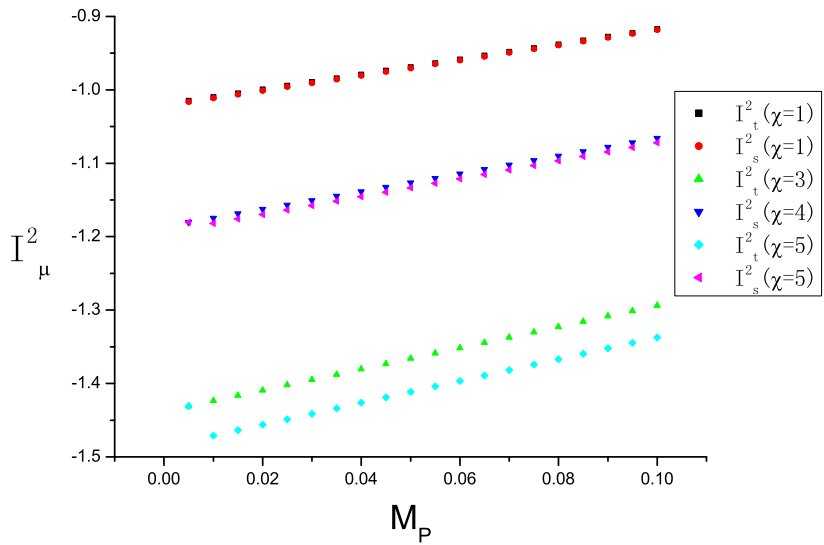}
 \end{center}
 \caption{The values of $I^{(2)}_\mu(m)$ as a function of $m_P$ are
 shown for $\chi=1$, $3$ and $5$. Here $I^{(1)}_t$ corresponds
 to $\mu=0$ and $I^{(1)}_s$ corresponds to $\mu=1,2,3$.
 Other bare parameters are:
 $\kappa_s=1$,  $M_0=-0.5$.
 \label{fig:I2}}
 \end{figure}
 As we pointed out above, if the quark mass is non-zero, quantities
 $I^{(1)}_\mu(m)$, $I^{(2)}_\mu(m)$ and $M^{(3)}(m)$ can
 be calculated directly using numerical integration
 once the bare parameters are given.
 In Fig.~\ref{fig:I1} and Fig.~\ref{fig:I2}, we have
 shown the values of $I^{(1)}_\mu$ and $I^{(2)}_\mu$ as functions
 of the propagator mass parameter $m_P$.
 Here the subscript ``t" corresponds
 to $\mu=0$ and the subscript ``s" corresponds to $\mu=1,2,3$.
 \begin{figure}[htb]
 \begin{center}
 \includegraphics[width=12.0cm,angle=0]{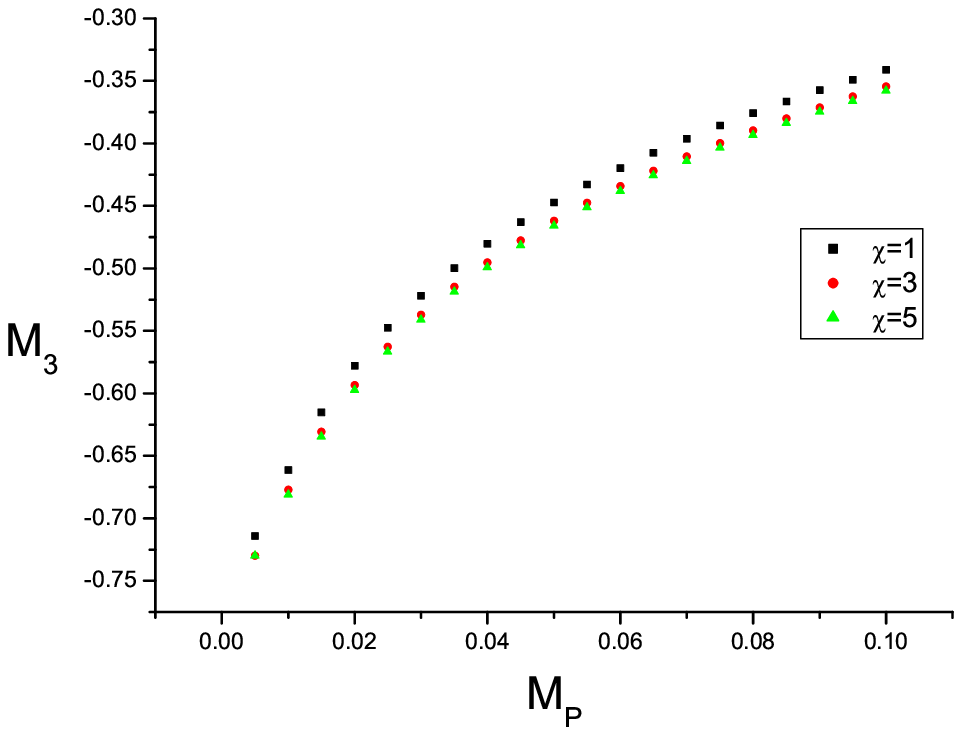}
 \end{center}
 \caption{The values of $M^{(3)}(m)$ as a function of $m_P$ are
 shown for $\chi=1$, $3$ and $5$.
 Other bare parameters are: $\kappa_s=1$,  $M_0=-0.5$.
 \label{fig:M3}}
 \end{figure}
 In Fig.~\ref{fig:M3}, we have
 shown the values of $M^{(3)}$  as a function
 of the propagator mass parameter $m_P$.

 In the massless case, the subtracted part of
 the loop integrals can be computed numerically
 following Eq.~(\ref{eq:Ifinite}) and Eq.~(\ref{eq:Mfinite}).
 Since the main purpose of this paper is to
 address massive overlap quarks, we will not
 list the numerical results for the massless case.

 Finally, our calculation can be easily translated into its
 tadpole-improved version following standard steps.
 For example, if we use the mean-field estimate (tree-level
 tadpole improved theory), Eq.~(\ref{eq:b_def}) is modified to:
 \be
 \tilde{b}(p)=\sum_\mu\kappa_\mu(1-u_\mu\cos p_\mu) +M_0
 \;,
 \ee
 where $u_\mu$ represents the mean-field value
 for the gauge field $U_\mu$. Note that for small
 lattice momenta, this amounts to a shift in parameter $M_0$:
 \be
 \tilde{M}_0=M_0+\sum_\mu\kappa_\mu(1-u_\mu)\;.
 \ee
 Similarly, the tuning of the hopping
 parameter $\kappa_t$ can also be discussed within the
 mean-field approximation. Note that these might be
 a rather good estimate for the correct value for
 the parameters in future Monte Carlo simulations.

 \section{Discussions and conclusions}

 In this paper, we have studied massive
 overlap fermions on anisotropic lattices.
 We argue that this setup can be useful in many
 lattice QCD studies. It is shown that, in order to restore the
 usual dispersion relation for the quark at
 small three-momenta, hopping parameter $\kappa_t$ has
 to be tuned according to the quark mass values.
 Quark propagator is calculated using bare
 perturbation theory to one-loop order.
 We find the wave-function and mass renormalization
 constants at various values of the bare parameters.
 These results serve as a guidance for the tuning
 of the parameters in real Monte Carlo simulations.

 \appendix
 \section{Vertex functions}
 \label{asec:vertex}

 Here we list the vertex functions used in the main text.
 The isotropic lattice counterparts for the vertex functions
 can be found in the literature~\cite{yamada98:overlap_perturb,yamada99:overlap_perturb,ishibashi00:overlap_perturb,fujikawa02:overlap_perturb}
 and the version for anisotropic lattice can be obtained accordingly
 after obvious modifications. The interaction vertex for
 the quark field, anti-quark field and one gluon field reads:
 \be
 \label{eq:vertex3}
 -g_0\frac{1-\frac{m}{2}}{\omega(q) +\omega(p)}T^A_{ab}\left\{
 V_{1\mu}\left(p + \frac{k}{2}\right) -
 \frac{X_0(q)}{\omega(q)}V_{1\mu}^\dagger\left(p +
  \frac{k}{2}\right)\frac{X_0(p)}{\omega(p)}\right\}.
 \ee
 The the two-gluon interaction vertex
 is much more complicated. It is found to be:
 \ba
 \label{eq:vertex4}
 &&  -g^2_0\left(1-\frac{m}{2}\right){1\over 2}\left\{T^A,T^B\right\}_{ab} \times
 \nonumber \\
 && \left[\frac{1}{\omega(q) + \omega(p)}\left\{ V_{2\mu}\left(p
  + \frac{k}{2}\right)\delta_{\mu\nu} -
  \frac{X_0(q)}{\omega(q)}V_{2\mu}^\dagger\left(p +
  \frac{k}{2}\right)\frac{X_0(p)}{\omega(p)} \delta_{\mu\nu}\right\}\right.\nonumber\\
 &&\;\;\;-\frac{1}{\{\omega(q) + \omega(p)\}\{\omega(p) + \omega(p+k_1)\}\{\omega(p+k_1) + \omega(q)\}}\nonumber\\
 && \times \left\{ X_0(q)V^\dagger_{1\mu}\left(p+k_2 +
  \frac{k_1}{2}\right)V_{1\nu}\left(p + \frac{k_2}{2}\right)\right.
  \nonumber \\
  &&+ V_{1\mu}\left(p+k_2 +
  \frac{k_1}{2}\right)X_0^\dagger(p+k_2) V_{1\nu}(p + \frac{k_2}{2})\nonumber\\
 &&  +  V_{1\mu}\left(p+k_2 + \frac{k_1}{2}\right)
  V_{1\nu}^\dagger\left(p + \frac{k_2}{2}\right)X_0(p)\nonumber \\
 &&  - \frac{\omega(q) +\omega(p) +\omega(p+k_2)}{\omega(q)\omega(p)\omega(p+k_2)}
 \times \nonumber \\
 && \left. X_0(q)V_{1\mu}^\dagger
 \left(p+k_2 + \frac{k_1}{2}\right)X_0(p+k_2)V_{1\nu}^\dagger\left(p + \frac{k_2}{2}\right)X_0(p)\right\}\nonumber \\
 &&\;\;\;-\frac{1}{\{\omega(q) + \omega(p)\}\{\omega(p) + \omega(p+k_1)\}\{\omega(p+k_1) + \omega(q)\}}\nonumber \\
 && \times \left\{ X_0(q)V^\dagger_{1\nu}\left(p+k_1 +
  \frac{k_2}{2}\right)V_{1\mu}\left(p + \frac{k_1}{2}\right)\right.
  \nonumber \\
 && + V_{1\nu}\left(p+k_1 + \frac{k_2}{2}\right)X_0^\dagger(p+k_1) V_{1\mu}\left(p + \frac{k_1}{2}\right)
 \nonumber \\
 &&  \hspace{0.5cm}+  V_{1\nu}\left(p+k_1 + \frac{k_2}{2}\right) V_{1\mu}^\dagger\left(p + \frac{k_1}{2}\right)X_0(p)\nonumber \\
 &&  \hspace{0.5cm}- \frac{\omega(q) +\omega(p) +\omega(p+k_1)}{\omega(q)\omega(p)\omega(p+k_1)}
 \times \nonumber \\
 && \left.\left.X_0(q)V_{1\nu}^\dagger
 \left(p+k_1 + \frac{k_2}{2}\right)X_0(p+k_1)V_{1\mu}^\dagger
 \left(p + \frac{k_1}{2}\right)X_0(p)\right\}\right]\;.
 \ea

 \bibliography{overlap}

\begin{thebibliography}{10}

\bibitem{kaplan92:DWF_idea}
D.~Kaplan.
\newblock A method for simulating chiral fermions on the lattice.
\newblock {\em Phys. Lett. B}, 288:342, 1992.

\bibitem{shamir93:DWF_boundary}
Y.~Shamir.
\newblock Chiral fermions from lattice boundaries.
\newblock {\em Nucl. Phys. B}, 406:90, 1993.

\bibitem{neuberger93:overlap_prl}
R.~Narayanan and H.~Neuberger.
\newblock Chiral fermions on the lattice.
\newblock {\em Phys. Rev. Lett.}, 71:3251, 1993.

\bibitem{neuberger94:overlap_npb}
R.~Narayanan and H.~Neuberger.
\newblock Chiral determinant as an overlap of two vacua.
\newblock {\em Nucl. Phys. B}, 412:574, 1994.

\bibitem{neuberger98:massless_vector}
Herbert Neuberger.
\newblock Vectorlike gauge theories with almost massless fermions on the
  lattice.
\newblock {\em Phys. Rev. D}, 57(9):5417--5433, May 1998.

\bibitem{neuberger98:massless_quark}
H.~Neuberger.
\newblock Exactly massless quarks on the lattice.
\newblock {\em Phys. Lett. B}, 417:141, 1998.

\bibitem{neuberger98:more_massless_quark}
H.~Neuberger.
\newblock More about exactly massless quarks on the lattice.
\newblock {\em Phys. Lett. B}, 427:353, 1998.

\bibitem{colin99}
C.~Morningstar and M.~Peardon.
\newblock The glueball spectrum from an anisotropic lattice study.
\newblock {\em Phys. Rev. D}, 60:034509, 1999.

\bibitem{chuan01:gluea}
C.~Liu.
\newblock A lattice study of the glueball spectrum.
\newblock {\em Chinese Physics Letter}, 18:187, 2001.

\bibitem{chen06:glueball}
Y.~Chen, A.~Alexandru, S.J. Dong, T.~Draper, I.~Horvath, F.X. Lee, K.F. Liu,
  N.~Mathur, C.~Morningstar, M.~Peardon, S.~Tamhankar, B.L. Young, and J.B.
  Zhang.
\newblock Glueball spectrum and matrix elements on anisotropic lattices.
\newblock {\em Phys. Rev. D}, 73:014516, 2006.

\bibitem{chen01:aniso}
P.~Chen.
\newblock Heavy quarks on anisotropic lattices: The charmonium spectrum.
\newblock {\em Phys. Rev. D}, 64:034509, 2001.

\bibitem{CPPACS02:aniso}
M.~Okamoto et~al.
\newblock Charmonium spectrum from quenched anisotropic lattice qcd.
\newblock {\em Phys. Rev. D}, 65:094508, 2002.

\bibitem{juettner03:Ds}
A.~Juettner and J.~Rolf.
\newblock A precise determination of the decay constant of the ds-meson in
  quenched qcd.
\newblock {\em Phys. Lett. B}, 560:59, 2003.

\bibitem{lewis01:aniso}
R.~Lewis, N.~Mathur, and R.M. Woloshyn.
\newblock Charmed baryons in lattice qcd.
\newblock {\em Phys. Rev. D}, 64:094509, 2001.

\bibitem{chuan02:pipiI2}
C.~Liu, J.~Zhang, Y.~Chen, and J.P. Ma.
\newblock Calculating the i=2 pion scattering length using tadpole improved
  clover wilson action on coarse anisotropic lattices.
\newblock {\em Nucl. Phys. B}, 624:360, 2002.

\bibitem{chuan04:KN}
G.~Meng, C.~Miao, X.~Du, and C.~Liu.
\newblock Lattice study on kaon nucleon scattering length in the i=1 channel.
\newblock {\em hep-lat/0309048}, 2003.

\bibitem{chuan04:Kpi}
C.~Miao, X.~Du, G.~Meng, and C.~Liu.
\newblock Lattice study on kaon pion scattering length in the $i=3/2$ channel.
\newblock {\em hep-lat/0403028}, 2004.

\bibitem{chuan04:pipi}
X.~Du, C.~Miao, G.~Meng, and C.~Liu.
\newblock $i=2$ pion scattering length with improved actions on anisotropic
  lattices.
\newblock {\em hep-lat/0404017}, 2004.

\bibitem{chuan06:DWF_perturb}
Xu~Feng, Xin Li, Wei Liu, and Chuan Liu.
\newblock Massive domain wall fermions on four-dimensional anisotropic
  lattices.
\newblock {\em JHEP}, 08:060, 2006.

\bibitem{colin97}
C.~Morningstar and M.~Peardon.
\newblock Efficient glueball simulations on anisotropic lattices.
\newblock {\em Phys. Rev. D}, 56:4043, 1997.

\bibitem{yamada98:overlap_perturb}
Atsushi Yamada.
\newblock Weak coupling expansion of a chiral gauge theory on a lattice in the
  overlap formulation.
\newblock {\em Nucl. Phys. B}, 514:399, 1998.

\bibitem{yamada99:overlap_perturb}
Y.~Kikukawa and A.~Yamada.
\newblock Weak coupling expansion of massless qcd with a ginsparg-wilson
  fermion and axial u(1) anomaly.
\newblock {\em Phys. Lett. B}, 448:265, 1999.

\bibitem{ishibashi00:overlap_perturb}
M.~Ishibashi, Y.~Kikukawa, T.~Noguchi, and A.~Yamada.
\newblock One-loop analyses of lattice qcd with the overlap dirac operator.
\newblock {\em Nucl. Phys. B}, 576:501, 2000.

\bibitem{fujikawa02:overlap_perturb}
Kazuo Fujikawa and Masato Ishibashi.
\newblock A perturbative study of a general class of lattice dirac operators.
\newblock {\em Phys. Rev. D}, 65:114504, 2002.

\bibitem{kfliu02:overlap_aniso}
K.F. Liu and S.J. Dong.
\newblock Heavy and light quarks with lattice chiral fermions.
\newblock {\em Int.J.Mod.Phys. A}, 20:7241, 2002.

\bibitem{ginsparg-wilson:1982}
P.H. Ginsparg and K.G. Wilson.
\newblock A remnant of chiral symmetry on the lattice.
\newblock {\em Phys. Rev.}, D25:2649, 1982.

\bibitem{luscher98:GWL_symmetry}
M.~L{\"u}scher.
\newblock Exact chiral symmetry on the lattice and the ginsparg-wilson
  relation.
\newblock {\em Phys. Lett. B}, 428:342, 1998.

\bibitem{hasenfratz98:GW_symmetry}
P.~Hasenfratz.
\newblock Lattice qcd without tuning, mixing and current renormalization.
\newblock {\em Nucl. Phys. B}, 525:401, 1998.

\bibitem{shigemitsu00:aniso_oneloop}
Stefan Groote and Junko Shigemitsu.
\newblock One-loop self energy and renormalization of the speed of light for
  some anisotropic improved quark actions.
\newblock {\em Phys. Rev. D}, 62:014508, 2000.

\end{thebibliography}
\end{document}